\documentclass[conference]{IEEEtran}
\IEEEoverridecommandlockouts
\usepackage{cite}
\usepackage{amsmath,amssymb,amsfonts}
\usepackage{algorithmic}
\usepackage{graphicx}
\usepackage{textcomp}
\usepackage{xcolor}
\usepackage{url}
\usepackage[export]{adjustbox}
\usepackage{paralist}

\DeclareMathSymbol{\mathdblquotechar}{\mathalpha}{letters}{`"}
\newcommand{\mathdblquote}{\mathtt{\mathdblquotechar}}
\begingroup\lccode`~=`"\lowercase{\endgroup
  \let~\mathdblquote
}
\AtBeginDocument{\mathcode`"="8000}

\def\graphwidth{0.8}

\def\cam#1{{#1}}

\def\BibTeX{{\rm B\kern-.05em{\sc i\kern-.025em b}\kern-.08em
    T\kern-.1667em\lower.7ex\hbox{E}\kern-.125emX}}
\begin{document}

\title{Lost in Disclosure: On The Inference of Password Composition Policies}

\author{\IEEEauthorblockN{Saul Johnson\IEEEauthorrefmark{1},
Jo\~ao F. Ferreira\IEEEauthorrefmark{2},
Alexandra Mendes\IEEEauthorrefmark{3} and Julien Cordry\IEEEauthorrefmark{1}}
\IEEEauthorblockA{\IEEEauthorrefmark{1}Software and Systems Research Group, Teesside University, Middlesbrough, UK \\
\IEEEauthorrefmark{2}Instituto Superior T\'ecnico, University of Lisbon and INESC-ID, Lisbon, Portugal \\
\IEEEauthorrefmark{3}HASLab, INESC TEC and Computer Science Department, University of Beira Interior, Covilh\~a, Portugal \\
Email: \IEEEauthorrefmark{1}\{saul.johnson,j.cordry\}@tees.ac.uk,
\IEEEauthorrefmark{2}joao@joaoff.com,
\IEEEauthorrefmark{3}alexandra@archimendes.com}}


\maketitle

\begin{abstract}
Large-scale password data breaches are becoming increasingly commonplace, which has enabled researchers to produce a substantial body of password security research utilising real-world password datasets, which often contain numbers of records in the tens or even hundreds of millions. While much study has been conducted on how password composition policies---sets of rules that a user must abide by when creating a password---influence the distribution of user-chosen passwords on a system, much less research has been done on inferring the password composition policy that a given set of user-chosen passwords was created under. In this paper, we state the problem with the naive approach to this challenge, and suggest a simple approach that produces more reliable results. We also present \textit{pol-infer}, a tool that implements this approach, and demonstrates its use in inferring password composition policies.
\end{abstract}


\section{Introduction}\label{sec:introduction}
When cybercriminals compromise a user credential database and release its contents into the public arena, a number of different interested parties might seek to obtain and use the data it contains, with varying goals in mind. These might include, for instance, other groups of cybercriminals seeking to employ the data in credential stuffing attacks \cite{alsaleh2012revisiting}, and security researchers seeking to understand user password choice on the system concerned \cite{weir2010testing,mazurek2013measuring,ur2016users}. In particular, the latter group may be concerned with the \textit{password composition policy} the passwords in the database were created under, in order to better understand how these rules around user password creation affect the distribution of user password choices.

Security researchers may find themselves confounded in this endeavour, however, because when the breached user credential database is released to the public, information about the password composition policy in place at the time of the breach is often not included. This could be because the party behind the breach does not think it relevant, wishes to keep their methods as secret as possible, or never sought this information out in the first place---after all, the password composition policy is of comparatively little interest to malicious actors seeking to directly employ the credentials in the database to criminal ends. The only other party known to have this information is the organisation that was the victim of the data breach in the first place, who by this point may be unable or unwilling to disclose any information regarding their security practices. Reasons for this might include, for example:

\begin{itemize}
    \item The organisation may have ceased to exist entirely, prior to the time at which the research in question is being conducted. There are several examples of this happening in the real world, for example the now-defunct Christian dating site \textit{singles.org} \cite{pauli2009exposed} which ceased to exist sometime after 2009 when their entire user credential database was compromised in plaintext.
    \item The organisation might be understandably reluctant to disclose any information regarding their security practices for fear of being further targeted or incriminating themselves by confessing to having taken inadequate measures to safeguard user data. This is especially the case in Europe, where tightening legislation around data protection \cite{gdpr2016} might make the latter point of particular concern. 
\end{itemize}

If we cannot obtain a description of the password composition policy from any of the organisations involved in the breach, this information has been \textit{lost in disclosure}---that is, lost somewhere in the process of the transfer of data between parties. We are therefore forced to turn to the data that we do have to attempt to infer as much of that lost information as we can.

\begin{table}[ht]
    \centering
    \caption{The four real-world breached password datasets studied in this work, alongside their corresponding policies according to \cite{golla2018accuracy,mayer2017second}, and numbers of passwords within them.}
    \label{tbl:dataset-specs}
    \begin{tabular}{|l|l|l|l|l|}
        \hline
        Dataset                                  & Policy                               & Size        \\ \hline
        RockYou \cite{cubrilovich2009rockyou}    & $length \geq 5$                      & 32,603,048  \\ \hline
        Yahoo \cite{gross2012yahoo}              & $length \geq 6$                      & 453,492     \\ \hline
        000webhost \cite{osborne2015000webhost}  & $length \geq 6 \wedge digits \geq 1$ & 15,271,208  \\ \hline
        LinkedIn \cite{burgess2016check}         & $length \geq 6$                      & 172,428,238 \\
        \hline
    \end{tabular}
\end{table}

There is no shortage of breached user credential databases available online. Arguably the most well-known of these, the RockYou set \cite{cubrilovich2009rockyou}, like many others (e.g. the Yahoo \cite{gross2012yahoo} or 000webhost \cite{osborne2015000webhost} sets) contains passwords that do not comply with the password composition policy in place when the breach happened (see Tables~\ref{tbl:dataset-specs} and \ref{tbl:compliance-data}). Reasons for this ``noise'' vary, but include:

\begin{itemize}
    \item Multiple password composition policies per dump---the RockYou set, for example, is an aggregate made up of at least two tables: one containing passwords to the main web application and one containing passwords used to log in to ``partner services'' (e.g. MySpace) which may enforce different policies \cite{cubrilovich2009rockyou}. Passwords created under old policies may also be present. RockYou, for instance, changed their policy after their data breach in 2009 from minimum 5 characters in length \cite{golla2018accuracy} to a stronger policy \cite{mayer2017second,florencio2010where}. \cam{In this case, our methodology gives the password composition policy that the majority of passwords were created under, though there is scope for improving upon this in future work (see Section~\ref{sec:conclusion}).}
    \item Formatting errors---when the raw data is being processed by the exfiltrating party, errors may be introduced if their data processing scripts are not robust. For example, passwords containing spaces may be read as two separate data points.
    \item Intentional padding---if cybercriminals initially offer the data for sale, the price that they are capable of obtaining is often contingent on the number of records it contains. It is therefore possible that the dataset may be intentionally padded with extra records, some of which might contain non-compliant passwords.
\end{itemize}

\begin{table}[ht]
    \centering
    \caption{A breakdown of the number of compliant and non-compliant passwords present in each dataset listed in Table~\ref{tbl:dataset-specs}, according to \cite{golla2018accuracy,mayer2017second}.}
    \label{tbl:compliance-data}
    \begin{tabular}{|l|l|l|l|l|}
        \hline
        Dataset                                  & Compliant  & Non-compliant    \\ \hline
        RockYou \cite{cubrilovich2009rockyou}    & 32524461   & 78587 (0.24\%)   \\ \hline
        Yahoo \cite{gross2012yahoo}              & 444942     & 8550 (1.89\%)    \\ \hline
        000webhost \cite{osborne2015000webhost}  & 14936872   & 334336 (2.19\%)  \\ \hline
        LinkedIn \cite{burgess2016check}         & 172409689  & 18549 (0.01\%)   \\
        \hline
    \end{tabular}
\end{table}

With ``noisy'' data like this, we cannot, for example, simply check for the shortest password in the database to determine the minimum password length constraint specified by the policy. In fact, the authors of one published work \cite{kelley2012guess} mention in their publication that the presence of ``non-password artifacts'' in the RockYou dataset factored in to their choice of research methods, at least in part due to the difficulty of filtering these out. This motivates us to search for a simple, easy-to-implement method to attempt to infer password composition policy rules from a password dataset, which would make filtering out at least some of these artifacts trivial. The remainder of this work outlines an alternative approach that we have found success with.

\paragraph{Contribution} We make the following concrete contributions in this work: 
\begin{inparaenum}[(i)]
\item for the first time, we draw attention to the problem of ``noise'' in publicly-available breached password datasets in the form of passwords that do not comply with the password composition policy in place when the breach occurred
\item we suggest an easy-to-implement approach to filtering out this noise by converting the problem to one of outlier detection, without consulting any organisation involved in the breach
\item we make \textit{pol-infer} \cite{johnson2019pol} available\footnote{Available for download at: \url{https://sr-lab.github.io/pol-infer/}}, the tool used to produce the data and visualisations in our results (Section~\ref{sec:results-real} and Section~\ref{sec:results-synth}). 
\end{inparaenum}

\paragraph{Outline} We have introduced and motivated the work in this Section~\ref{sec:introduction}. We describe related work in Section~\ref{sec:related}. In Section~\ref{sec:methodology} we describe our approach in detail, showing the results we are able to obtain from the four password datasets shown in Table~\ref{tbl:compliance-data} in Section~\ref{sec:results-real}. In Section~\ref{sec:results-synth} we apply our methodology to datasets created to simulate both intentional padding and processing with error-prone data processing scripts. We conclude in Section~\ref{sec:conclusion}, discussing the limitations of our approach and potential future work.

\section{Related Work}\label{sec:related}
\cam{We are not aware of any existing published work that explores the automation of password composition policy inference from large datasets. Previous research has involved determining the password composition policies used by active services. A study by Flor\^encio and Herley \cite{florencio2010where} gathered password composition policy information by creating an account on the service, where possible, and performing web searches otherwise. This study was later replicated by Mayer et al. in \cite{mayer2017second}. In \cite{golla2018accuracy}, Golla and D\"urmuth make extensive use of password data dumps where the password composition policy is known.}

\section{Methodology}\label{sec:methodology}
Our approach is applicable to any numerically-typed password attribute $\alpha$ which is a function of type $Password \rightarrow \mathbb{N}$ which extracts some password property (e.g. length). By default, \textit{pol-infer} supports the password attributes in Table~\ref{tbl:attributes}, sufficient to capture the policies used in the study by Shay et al. \cite{shay2016designing} with the exception of the dictionary check on the \textit{comprehensive8} policy, which cannot be expressed as an attribute of this type.

\begin{table}[ht]
    \centering
    \caption{Password attributes usable with \textit{pol-infer} by default. Any attribute appearing the table below can be used by the tool to infer password composition policies.}
    \label{tbl:attributes}
    \begin{tabular}{|p{2cm}|p{5cm}|}
        \hline
        Attribute ($\alpha$) & Description                                                 \\ \hline
        length               & The number of characters in the password (i.e. its length). \\ \hline
        words                & The number of words in the password. We define ``words'' 
                               in the same way as in \cite{shay2016designing}---as 
                               ``letter sequences separated by a nonletter sequence''.     \\ \hline
        lowers               & The number of lowercase letters in the password.            \\ \hline
        uppers               & The number of uppercase letters in the password.            \\ \hline
        digits               & The number of digits in the password.                       \\ \hline
        symbols              & The number of non-alphanumeric characters in the password.  \\ \hline
        classes              & The number of character classes in the password. We 
                               recognise four character classes in the popular LUDS 
                               scheme---lowercase, uppercase, digits and symbols.          \\ \hline
    \end{tabular}
\end{table}

For instance, let us suppose we wish to infer the minimum length constraint specified by the policy that the 000webhost set \cite{osborne2015000webhost} was created under (that is, $\alpha = length$). In this case, previous research \cite{golla2018accuracy} has established that the answer is $6$, and yet the data in Table~\ref{tbl:example-data} would seem to contradict this---there are passwords shorter than this present in the data.

\begin{table}[ht]
    \centering
    \caption{Frequencies $f(l)$ of passwords of different lengths $l$ in the 000webhost set \cite{osborne2015000webhost}, alongside their cumulative frequencies $cum(l)$ and the multiplier $mult(l)$ required to reach the cumulative frequency of the next length $cum(l+1)$.}
    \label{tbl:example-data}
    \begin{tabular}{|l|l|l|l|}
        \hline
        $l$ & $f(l)$  & $cum(l)$ & $mult(l)$ \\ \hline
        1   & 306     & 306      & 6.03   \\ \hline 
        2   & 1540    & 1846     & 1.42   \\ \hline 
        3   & 775     & 2621     & 1.47   \\ \hline 
        4   & 1221    & 3842     & 1.66   \\ \hline 
        5   & 2456    & 6388     & 137.23 \\ \hline 
        6   & 870209  & 876597   & 2.38   \\ \hline 
        7   & 1208092 & 2084689  & ---    \\ \hline
    \end{tabular}
\end{table}

It is readily apparent how the data in Table~\ref{tbl:example-data} may be used to determine the minimum length constraint in the 000webhost policy. By observing the outlying value of $137.23$ in the $mult(l)$ column, we can see that we now have an outlier detection problem. In Table~\ref{tbl:example-data}, for every length $l$:

\begin{equation*}
    mult(l) = \frac{cum(l + 1)}{cum(l)}
\end{equation*}

We can infer the minimum password length enforced by the password composition policy under which this data was created by looking for the outlying ``sudden increase'' in $f(l)$, taking $l + 1$ where:

\begin{equation*}
    mult(l) = max(\{mult(m) | m \in \mathbb{N}\})
\end{equation*}

For the 000webhost data, this gives us the correct answer $6$. By examining the number of digits in a password, as opposed to password length (that is to say $\alpha = digits$), we are also able to determine that the 000webhost policy demands that passwords contain at least one digit (see Section~\ref{sec:results-real}). 

By setting a lower threshold on $mult(\alpha)$ we are able to specify a cutoff point $c$ below which we assume there is no constraint in place on the attribute $\alpha$ in question. For $\alpha \in \{length, digits, uppers\}$, we have found success using a value of $2$ as this threshold (i.e. $c=2$). For example, consider that the 000webhost policy does not demand that any uppercase letters be present in passwords.

\begin{table}[ht]
    \centering
    \caption{\cam{Frequencies $f(u)$, cumulative frequencies $cum(u)$ and multipliers $mult(u)$ of passwords containing different numbers of uppercase letters $u$ in the 000webhost set \cite{osborne2015000webhost}.}}
    \label{tbl:000webhsot-uppercase-data}
    \begin{tabular}{|l|l|l|l|}
        \hline
        $u$ & $f(u)$   & $cum(u)$ & $mult(u)$ \\ \hline
        0   & 12366006 & 12366006 & 1.08      \\ \hline 
        1   & 1049727  & 13415733 & 1.02      \\ \hline 
        2   & 315637   & 13731370 & 1.02      \\ \hline 
        3   & 267042   & 13998412 & 1.02      \\ \hline 
        4   & 260061   & 14258473 & 1.02      \\ \hline 
        5   & 241305   & 14499778 & 1.02      \\ \hline 
        6   & 220202   & 14719980 & 1.01      \\ \hline 
        7   & 187806   & 14907786 & ---       \\ \hline
    \end{tabular}
\end{table}

As no value in Table~\ref{tbl:000webhsot-uppercase-data} is outlying above the default cutoff point of $2$, we conclude that there was likely no constraint on minimum number of uppercase letters present in the password policy when the dataset was created.

\section{Results: Real Data}\label{sec:results-real}
We present a set of results demonstrating the success of our approach when used to infer minimum password length specified by the policy under which 4 different data sets were created.
 
\begin{itemize}
    \item \textbf{RockYou}---breached in plaintext from an online gaming service of the same name circa 2009 \cite{cubrilovich2009rockyou}. The policy in place at the time enforced a minimum length of 5 characters, with no other constraints \cite{golla2018accuracy}. Contains a total of 32,603,048 passwords.
    \item \textbf{Yahoo}---breached from the Yahoo Voices online publishing platform circa 2012 \cite{gross2012yahoo}. The policy in place at the time of the breach enforced a minimum length of 6 characters with no other requirements \cite{mayer2017second}. Contains 453,492 passwords.
    \item \textbf{000webhost}---breached from the web hosting service of the same name circa 2015 \cite{osborne2015000webhost}. The policy in place at the time of the breach enforced a minimum length of 6 characters, with at least one numeric digit \cite{golla2018accuracy}. Contains 15,271,208 passwords.
    \item \textbf{LinkedIn}---breached from the professional social networking site of the same name circa 2012, the true extent of this breach was uncovered in 2016 as much bigger than was initially made public \cite{burgess2016check}. Unsalted password hashes in SHA-1 format were extracted, of which $\approx98\%$ have since been cracked. It is these cracked passwords we use in this work. The policy in place at the time of the breach enforced a minimum length of 6 characters with no other requirements \cite{mayer2017second}. Contains 172,428,238 passwords.
\end{itemize}

The results that follow were produced using \textit{pol-infer}---a tool we make available \cite{johnson2019pol} for inferring password composition policies from large datasets using the approach we describe in Section~\ref{sec:methodology}.

\subsection{The RockYou Set (2009)}
Previous research has established that the majority of the RockYou set \cite{cubrilovich2009rockyou} was created under a policy enforcing minimum length $5$ with no other requirements \cite{golla2018accuracy}.

\begin{figure}[ht]
    \centering
    \includegraphics[width=\graphwidth\columnwidth]{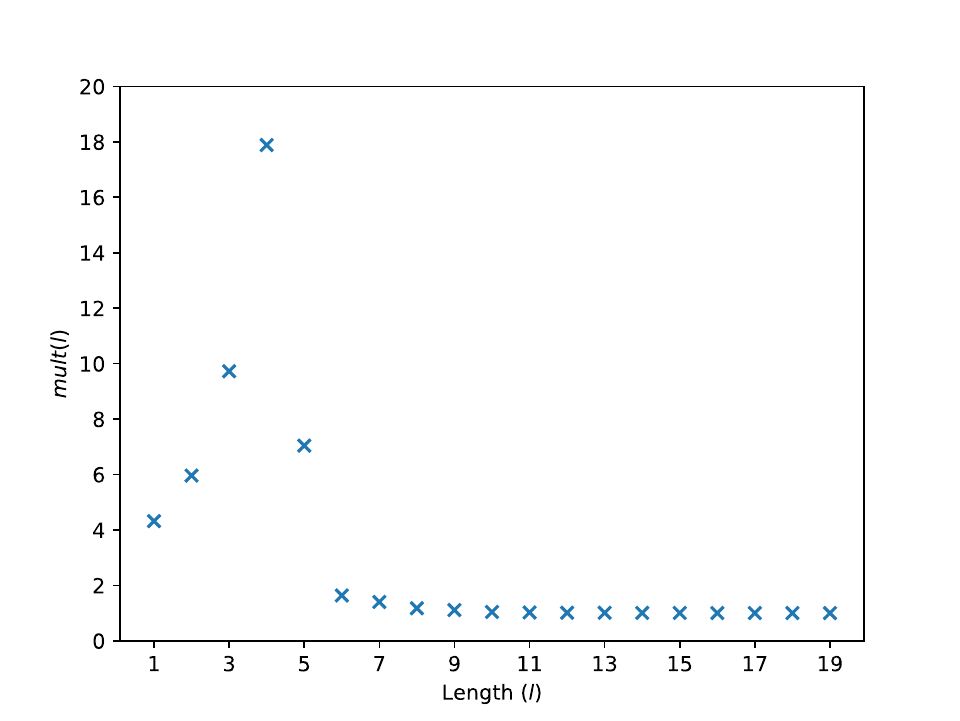}
    \caption{\cam{Passwords of different lengths $l$ in the RockYou set \cite{cubrilovich2009rockyou}, plotted against $mult(l)$.}}
    \label{fig:rockyou-length}
\end{figure}

The outlying point at $l = 4$ in Figure~\ref{fig:rockyou-length} indicates that the password composition policy that most of the passwords in the set were created under enforces a minimum length of $5$. This aligns with existing literature \cite{golla2018accuracy}.
    
\subsection{The Yahoo Set (2012)}
Previous research has established that the majority of the Yahoo set \cite{gross2012yahoo} was created under a policy enforcing minimum length $6$ with no other requirements \cite{mayer2017second}.

\begin{figure}[ht]
    \centering
    \includegraphics[width=\graphwidth\columnwidth]{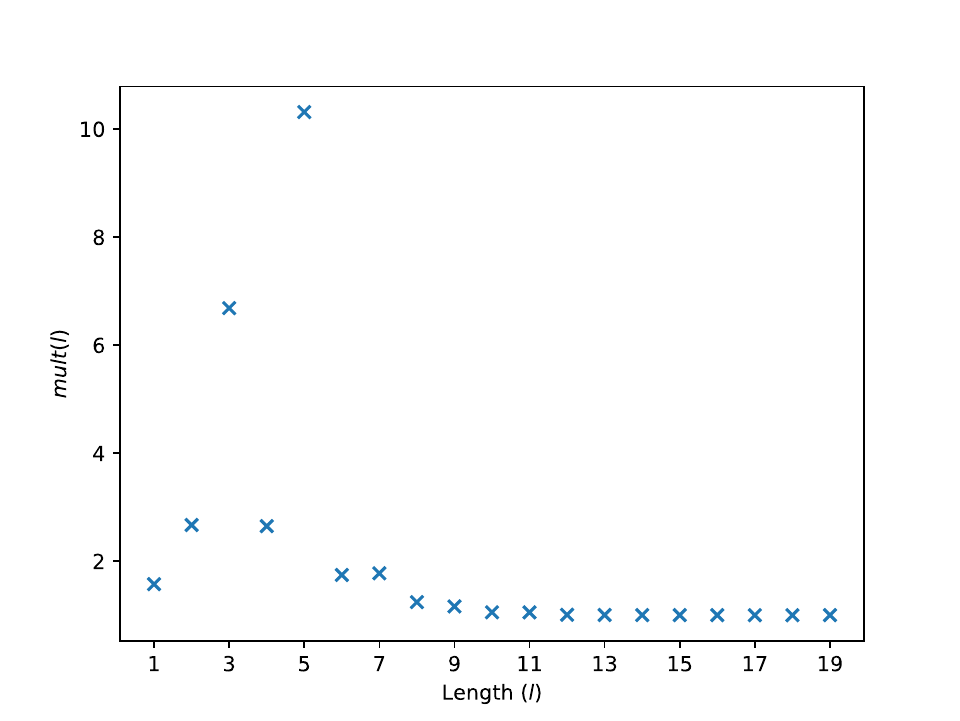}
    \caption{\cam{Passwords of different lengths $l$ in the Yahoo set \cite{gross2012yahoo}, plotted against $mult(l)$.}}
    \label{fig:yahoo-length}
\end{figure} 

The outlying point at $l = 5$ in Figure~\ref{fig:yahoo-length} indicates that the password composition policy that most of the passwords in the set were created under enforces a minimum length of $6$. This aligns with existing literature \cite{mayer2017second}.

\begin{figure}[ht]
    \centering
    \includegraphics[width=\graphwidth\columnwidth]{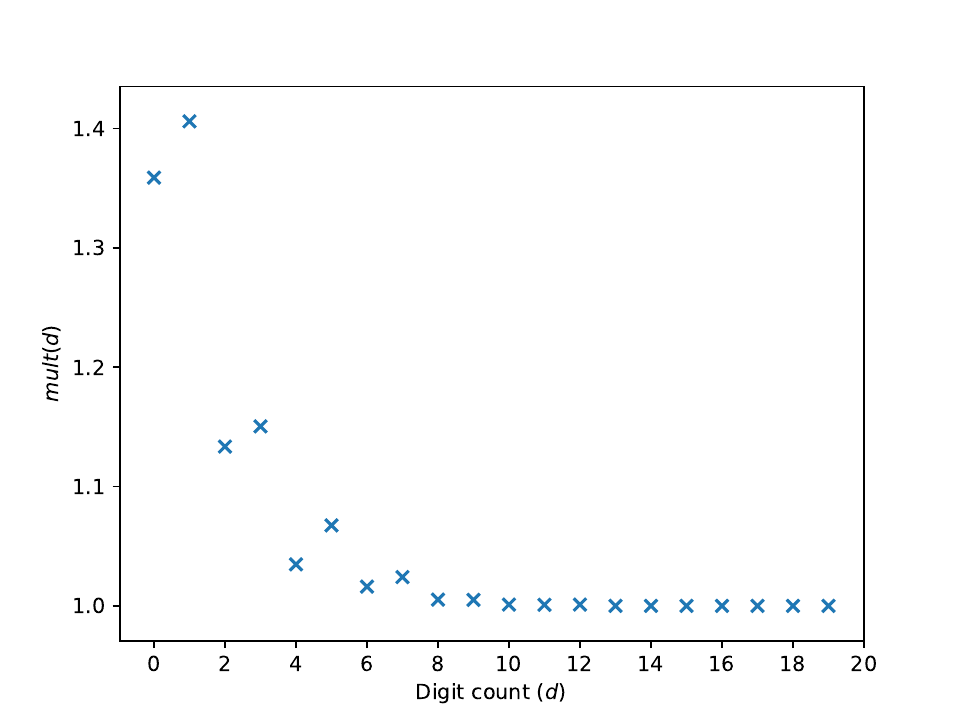}
    \caption{\cam{Passwords containing different numbers of digits $d$ in the Yahoo set \cite{gross2012yahoo}, plotted against $mult(d)$.}}
    \label{fig:yahoo-digits}
\end{figure} 

\subsubsection{Inferring the Absence of Constraints}
As no points in Figure~\ref{fig:yahoo-digits} are present above the default \textit{pol-infer} \cite{johnson2019pol} cutoff point of $c=2$, the tool indicates that there was likely no constraint on minimum number of digits present in the password policy when the Yahoo dataset was created. This aligns with existing literature \cite{mayer2017second}.

\subsection{The 000webhost Set (2015)}
Previous research has established that the majority of the 000webhost set \cite{osborne2015000webhost} was created under a policy enforcing minimum length $6$ with the additional requirement that passwords must contain at least one digit \cite{golla2018accuracy}.

\begin{figure}[ht]
    \centering
    \includegraphics[width=\graphwidth\columnwidth]{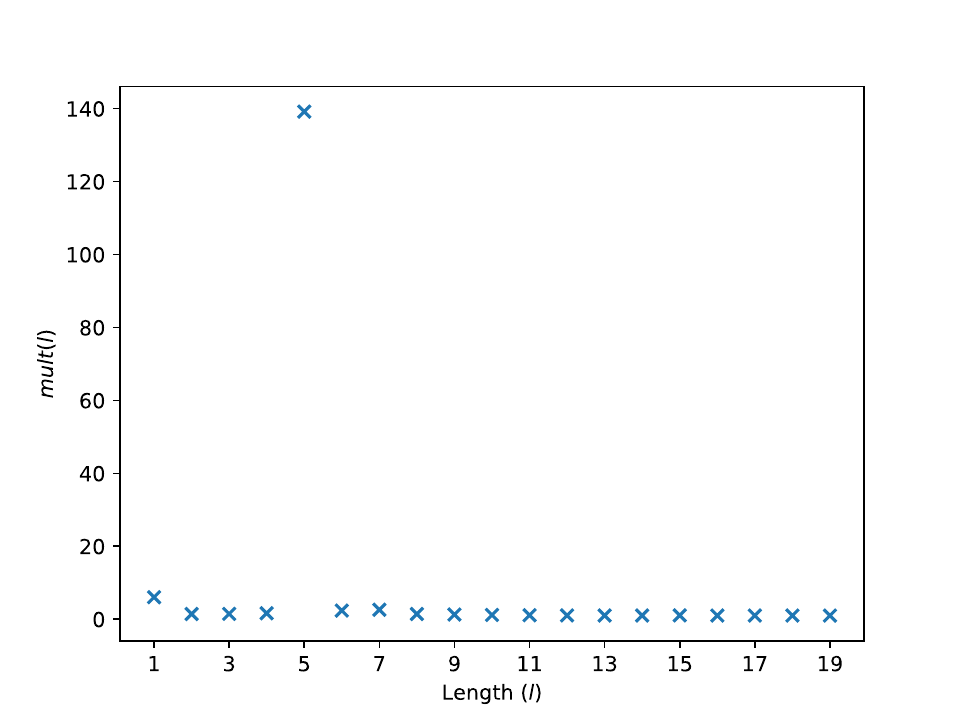}
    \caption{\cam{Passwords of different lengths $l$ in the 000webhost set \cite{osborne2015000webhost}, plotted against $mult(l)$.}}
    \label{fig:000webhost-length}
\end{figure}

The outlying point at $l = 5$ in Figure~\ref{fig:000webhost-length} indicates that the password composition policy that most of the passwords in the set were created under enforces a minimum length of $6$. This aligns with existing literature \cite{golla2018accuracy}.

\begin{figure}[ht]
    \centering
    \includegraphics[width=\graphwidth\columnwidth]{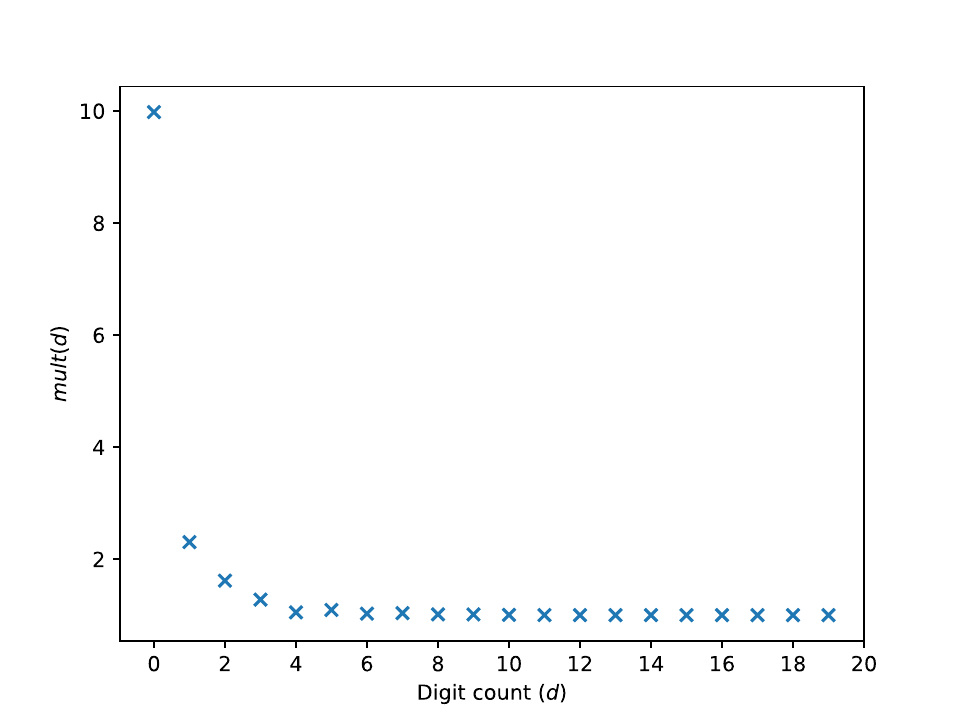}
    \caption{\cam{Passwords containing different numbers of digits $d$ in the 000webhost set \cite{osborne2015000webhost}, plotted against $mult(d)$.}}
    \label{fig:000webhost-digits}
\end{figure} 

The outlying point at $d = 0$ in Figure~\ref{fig:000webhost-digits} indicates that the password composition policy that most of the passwords in the set were created under enforces a minimum of $1$ digit in passwords.

\subsection{The LinkedIn Set (2016)}
Previous research has established that the majority of the LinkedIn set \cite{burgess2016check} was created under a policy enforcing minimum length $6$ with no other requirements \cite{golla2018accuracy}.

\begin{figure}[ht]
    \centering
    \includegraphics[width=\graphwidth\columnwidth]{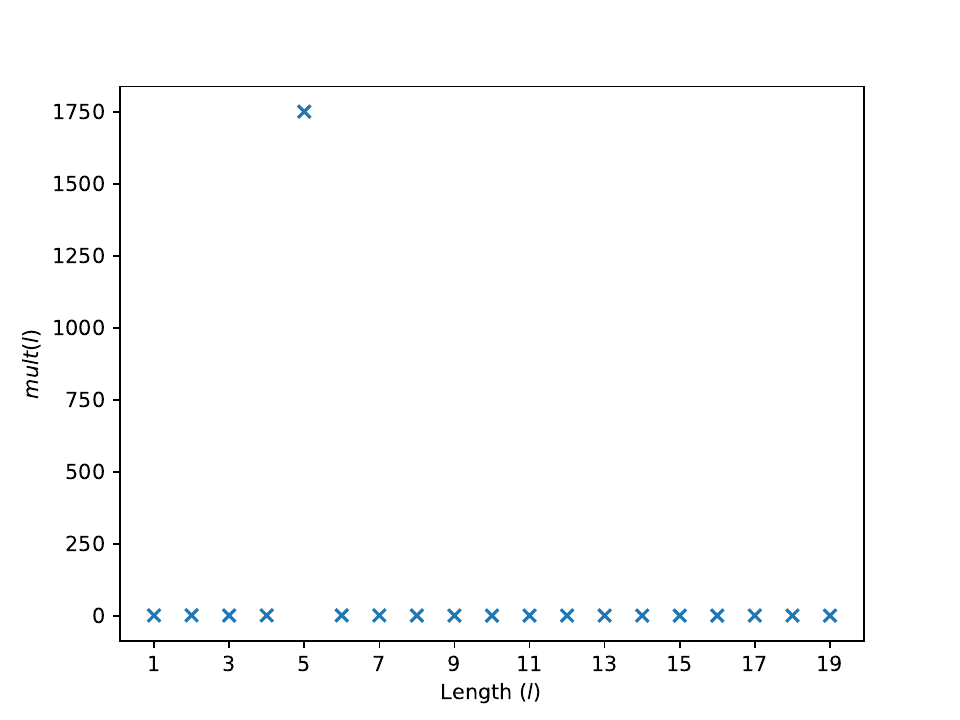}
    \caption{\cam{Passwords of different lengths $l$ in the LinkedIn set \cite{burgess2016check}, plotted against $mult(l)$.}}
    \label{fig:linkedin-length}
\end{figure} 

The outlying point at $l = 5$ in Figure~\ref{fig:linkedin-length} indicates that the password composition policy that most of the passwords in the set were created under enforces a minimum length of $6$. This aligns with existing literature \cite{golla2018accuracy}.

\section{Results: Synthetic Data}\label{sec:results-synth}
In order to simulate the effect of some of the circumstances mentioned in Section~\ref{sec:introduction} that could potentially create non-compliant ``noise'' in real-world password datasets, we created the following synthetic datasets:

\begin{itemize}
    \item \textbf{2word12\_linkedin\_padded}---The LinkedIn dataset \cite{burgess2016check} filtered according to a 2word12 policy (at least 12 characters long, at least 2 letter sequences separated by a non-letter sequence) to leave 1,511,786 passwords. This has then been combined with the \textit{singles.org} dataset \cite{pauli2009exposed} (16,248 passwords), \textit{elitehacker} dataset (1000 passwords), \textit{hak5} dataset \cite{constantin2009security} (2987 passwords), and \textit{faithwriters} dataset \cite{greenberg2010researcher} (9709 passwords). This is designed to simulate intentional padding of a dataset created under one policy with several other smaller datasets in order to increase its resale value.
    \item \textbf{2class8\_linkedin\_errors}---The LinkedIn dataset \cite{burgess2016check} filtered according to 2class8 policy (at least 8 characters long, at least 2 character classes present from lowercase, uppercase, digits and symbols) to leave 65,271,156 passwords. For every password in this dataset containing either a space or a comma, this password has then been split into two or more separate strings along these tokens, leading to the creation of 404,547 additional records. This simulates the type of formatting error that might be introduced by processing scripts after the dataset has been exfiltrated.
\end{itemize}

\subsection{Intentional Padding}
\begin{figure}[ht]
    \centering
    \includegraphics[width=\graphwidth\columnwidth]{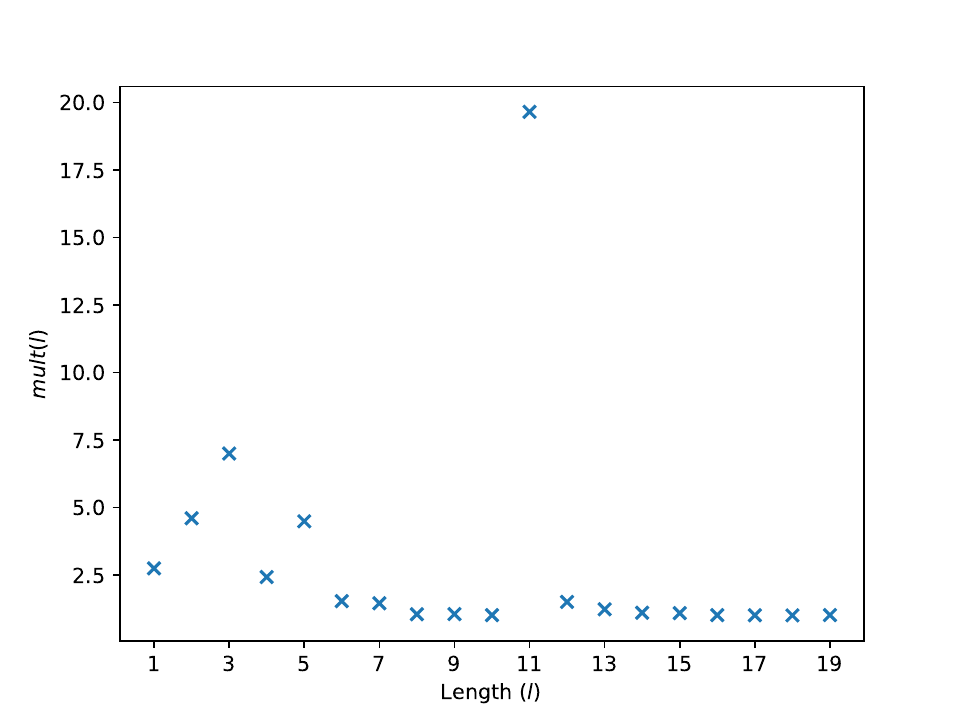}
    \caption{Passwords of different lengths $l$ in the 2word12\_linkedin\_padded synthetic dataset, plotted against the multiplier $mult(l)$ required to reach the cumulative frequency of the next length $cum(l+1)$.}
    \label{fig:synth-padded-length}
\end{figure}

Figure~\ref{fig:synth-padded-length} and Table~\ref{tbl:synth-padded-words} show the use of our methodology to recover the original password composition policy of 2word12\_linkedin\_padded (2word12). The outlying points at $l=11$ and $w=1$ give us a length and word count of $12$ and $2$ respectively.

\begin{table}[ht]
    \centering
    \caption{\cam{Frequencies $f(w)$, cumulative frequencies $cum(w)$ and multipliers $mult(w)$ of passwords containing different numbers of words $w$ in the 2word12\_linkedin\_padded synthetic dataset.}}
    \label{tbl:synth-padded-words}
    \begin{tabular}{|l|l|l|l|}
    \hline
        $w$ & $f(w)$  & $cum(w)$  & $mult(w)$ \\ \hline
        0   & 2500    & 2500      & 11.18     \\ \hline 
        1   & 25460   & 27960     & 39.39     \\ \hline 
        2   & 1073513 & 1101473   & 1.17      \\ \hline 
        3   & 190996  & 1292469   & 1.07      \\ \hline 
        4   & 89916   & 1382385   & ---       \\ \hline
    \end{tabular}
\end{table}

\subsection{Formatting Errors}
\begin{figure}[ht]
    \centering
    \includegraphics[width=\graphwidth\columnwidth]{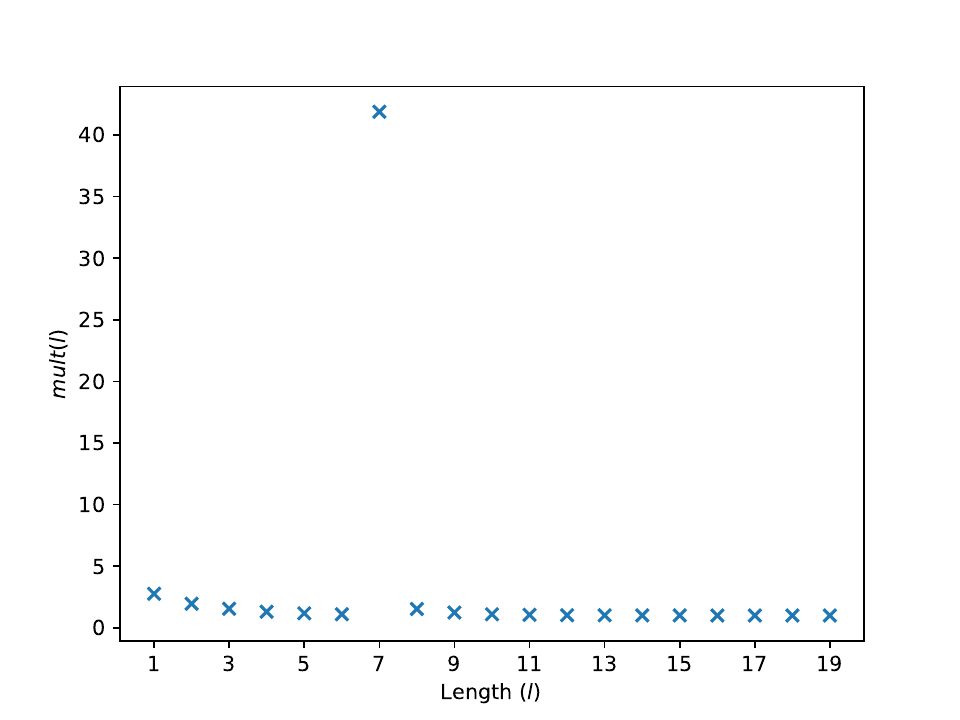}
    \caption{Passwords of different lengths $l$ in the 2word12\_linkedin\_errors synthetic dataset, plotted against $mult(l)$.}
    \label{fig:synth-errors-length}
\end{figure}

Figure~\ref{fig:synth-errors-length} and Table~\ref{tbl:synth-errors-classes} show the use of our methodology to recover the original password composition policy of 2class8\_linkedin\_errors (2class8). The outlying points at $l=7$ and $c=1$ give us a length and class count of $8$ and $2$ respectively.

\begin{table}[ht]
    \centering
    \caption{\cam{Frequencies $f(c)$, cumulative frequencies $cum(c)$ and multipliers $mult(c)$ of passwords containing different numbers of words $c$ in the 2word12\_linkedin\_errors synthetic dataset.}}
    \label{tbl:synth-errors-classes}
        \begin{tabular}{|l|l|l|l|}
        \hline
        $c$ & $f(c)$   & $cum(c)$  & $mult(c)$ \\ \hline
        1   & 591820   & 591820    & 84.87     \\ \hline 
        2   & 49637360 & 50229180  & 1.27      \\ \hline 
        3   & 13401629 & 63630809  & 1.03      \\ \hline 
        4   & 2044894  & 65675703  & ---       \\ \hline
    \end{tabular}
\end{table}

\section{Conclusion}\label{sec:conclusion}
In this work, we have demonstrated a simple, easy-to-implement methodology for inferring the password composition policy under which a password data dump was created without the need to interact with any of the parties involved in its disclosure. Once we have done this, we are able to trivially filter out non-compliant passwords if we so wish. We make \textit{pol-infer}, the tool implementing this methodology that we used to produce the results in Sections~\ref{sec:results-real} and \ref{sec:results-synth}, freely available \cite{johnson2019pol}. We show that results obtained by this tool agree with existing literature on several real-world password datasets, and that it is effective on datasets generated to mimic those that might arise as a result of intentional padding or buggy data processing.

\paragraph{Limitations} While our approach is capable of approximately inferring password composition policies that place constraints on specific password attributes, it cannot offer a guarantee that the inferred policy is accurate or complete. As an example of a password composition policy rule that would be very difficult to infer, consider a rule that limits password length to a maximum of 1024 characters. As very few user-chosen passwords would be in violation of this rule even in its absence, its impact on user password choice would be very limited, making its inference very difficult. 

\paragraph{Future work} \cam{Where time and date of account creation is available in password data dumps, it may be possible to detect with some accuracy the date and time of any password composition policy changes, offering new insight into the organisation's internal security practices. This may require \textit{pol-infer} to become more modular, acting as a framework capable of hosting different inference algorithms.}
Work on \textit{pol-infer} is planned to make policy inference more automated \cam{and comprehensive (e.g. inference of dictionary checks),} with an option to generate password composition policy names in the style used by \cite{shay2016designing}. \cam{We plan to make use of \textit{pol-infer} and the methodology we propose in this work to help prepare password data for use in research into other aspects of password security, such as formally verified password composition policy enforcement software \cite{ferreira2017certified}.}

\raggedright 

\bibliographystyle{IEEEtran}
\bibliography{main}

\begin{thebibliography}{10}
\providecommand{\url}[1]{#1}
\csname url@samestyle\endcsname
\providecommand{\newblock}{\relax}
\providecommand{\bibinfo}[2]{#2}
\providecommand{\BIBentrySTDinterwordspacing}{\spaceskip=0pt\relax}
\providecommand{\BIBentryALTinterwordstretchfactor}{4}
\providecommand{\BIBentryALTinterwordspacing}{\spaceskip=\fontdimen2\font plus
\BIBentryALTinterwordstretchfactor\fontdimen3\font minus
  \fontdimen4\font\relax}
\providecommand{\BIBforeignlanguage}[2]{{%
\expandafter\ifx\csname l@#1\endcsname\relax
\typeout{** WARNING: IEEEtran.bst: No hyphenation pattern has been}%
\typeout{** loaded for the language `#1'. Using the pattern for}%
\typeout{** the default language instead.}%
\else
\language=\csname l@#1\endcsname
\fi
#2}}
\providecommand{\BIBdecl}{\relax}
\BIBdecl

\bibitem{alsaleh2012revisiting}
M.~{Alsaleh}, M.~{Mannan}, and P.~C. {van Oorschot}, ``Revisiting defenses
  against large-scale online password guessing attacks,'' \emph{IEEE
  Transactions on Dependable and Secure Computing}, vol.~9, no.~1, pp.
  128--141, Jan 2012.

\bibitem{weir2010testing}
M.~Weir, S.~Aggarwal, M.~Collins, and H.~Stern, ``Testing metrics for password
  creation policies by attacking large sets of revealed passwords,'' in
  \emph{Proceedings of the 17th ACM Conference on Computer and Communications
  Security}, ser. CCS '10.\hskip 1em plus 0.5em minus 0.4em\relax New York, NY,
  USA: ACM, 2010, pp. 162--175.

\bibitem{mazurek2013measuring}
M.~L. Mazurek, S.~Komanduri, T.~Vidas, L.~Bauer, N.~Christin, L.~F. Cranor,
  P.~G. Kelley, R.~Shay, and B.~Ur, ``Measuring password guessability for an
  entire university,'' in \emph{Proceedings of the 2013 ACM SIGSAC Conference
  on Computer \&\#38; Communications Security}, ser. CCS '13.\hskip 1em plus
  0.5em minus 0.4em\relax New York, NY, USA: ACM, 2013, pp. 173--186.

\bibitem{ur2016users}
B.~Ur, J.~Bees, S.~M. Segreti, L.~Bauer, N.~Christin, and L.~F. Cranor, ``Do
  users' perceptions of password security match reality?'' in \emph{Proceedings
  of the 2016 CHI Conference on Human Factors in Computing Systems}, ser. CHI
  '16.\hskip 1em plus 0.5em minus 0.4em\relax New York, NY, USA: ACM, 2016, pp.
  3748--3760.

\bibitem{pauli2009exposed}
D.~Pauli, ``Exposed web site a reminder for use of multiple passwords | network
  world,''
  \url{https://www.networkworld.com/article/2263760/exposed-web-site-a-reminder-for-use-of-multiple-passwords.html},
  Feb 2009, (Accessed on 07/25/2019).

\bibitem{gdpr2016}
{European Parliament}, ``{Regulation (EU) 2016/679 of the European Parliament
  and of the Council of 27 April 2016 on the protection of natural persons with
  regard to the processing of personal data and on the free movement of such
  data, and repealing Directive 95/46/EC (General Data Protection
  Regulation)},'' \emph{Official Journal of the European Union}, vol.~59, pp.
  1--88, 2016.

\bibitem{golla2018accuracy}
M.~Golla and M.~D\"{u}rmuth, ``On the accuracy of password strength meters,''
  in \emph{Proceedings of the 2018 ACM SIGSAC Conference on Computer and
  Communications Security}, ser. CCS '18.\hskip 1em plus 0.5em minus
  0.4em\relax New York, NY, USA: ACM, 2018, pp. 1567--1582.

\bibitem{mayer2017second}
P.~Mayer, J.~Kirchner, and M.~Volkamer, ``A second look at password composition
  policies in the wild: Comparing samples from 2010 and 2016,'' in
  \emph{Thirteenth Symposium on Usable Privacy and Security ({SOUPS}
  2017)}.\hskip 1em plus 0.5em minus 0.4em\relax Santa Clara, CA: {USENIX}
  Association, 2017, pp. 13--28.

\bibitem{cubrilovich2009rockyou}
N.~Cubrilovich, ``Rockyou hack: From bad to worse | techcrunch,''
  \url{https://techcrunch.com/2009/12/14/rockyou-hack-security-myspace-facebook-passwords/},
  Dec 2009, (Accessed on 04/10/2019).

\bibitem{gross2012yahoo}
D.~Gross, ``Yahoo hacked, 450,000 passwords posted online - cnn,''
  \url{https://edition.cnn.com/2012/07/12/tech/web/yahoo-users-hacked}, Jul
  2012, (Accessed on 04/10/2019).

\bibitem{osborne2015000webhost}
C.~Osborne, ``000webhost hacked, 13 million customers exposed | zdnet,''
  \url{https://www.zdnet.com/article/000webhost-hacked-13-million-customers-exposed/},
  Oct 2015, (Accessed on 04/10/2019).

\bibitem{burgess2016check}
M.~Burgess, ``Check if your linkedin account was hacked | wired uk,''
  \url{https://www.wired.co.uk/article/linkedin-data-breach-find-out-included},
  May 2016, (Accessed on 07/26/2019).

\bibitem{florencio2010where}
D.~Flor\^{e}ncio and C.~Herley, ``Where do security policies come from?'' in
  \emph{Proceedings of the Sixth Symposium on Usable Privacy and Security},
  ser. SOUPS '10.\hskip 1em plus 0.5em minus 0.4em\relax New York, NY, USA:
  ACM, 2010, pp. 10:1--10:14.

\bibitem{kelley2012guess}
P.~G. {Kelley}, S.~{Komanduri}, M.~L. {Mazurek}, R.~{Shay}, T.~{Vidas},
  L.~{Bauer}, N.~{Christin}, L.~F. {Cranor}, and J.~{Lopez}, ``Guess again (and
  again and again): Measuring password strength by simulating password-cracking
  algorithms,'' in \emph{2012 IEEE Symposium on Security and Privacy}, May
  2012, pp. 523--537.

\bibitem{johnson2019pol}
S.~Johnson, ``sr-lab/pol-infer: Inferring password composition policies from
  breached user credential databases.''
  \url{https://github.com/sr-lab/pol-infer}, 4 2019, (Accessed on 04/12/2019).

\bibitem{shay2016designing}
R.~Shay, S.~Komanduri, A.~L. Durity, P.~S. Huh, M.~L. Mazurek, S.~M. Segreti,
  B.~Ur, L.~Bauer, N.~Christin, and L.~F. Cranor, ``Designing password policies
  for strength and usability,'' \emph{ACM Trans. Inf. Syst. Secur.}, vol.~18,
  no.~4, pp. 13:1--13:34, May 2016.

\bibitem{constantin2009security}
L.~Constantin, ``Security gurus 0wned by black hats,''
  \url{https://news.softpedia.com/news/Security-Gurus-0wned-by-Black-Hats-117934.shtml},
  Jul 2009, (Accessed on 05/10/2019).

\bibitem{greenberg2010researcher}
A.~Greenberg, ``Researcher creates clearinghouse of 14 million hacked
  passwords,''
  \url{https://www.forbes.com/sites/andygreenberg/2010/08/26/researcher-creates-clearinghouse-of-14-million-hacked-passwords/#7bacb64318fd},
  Aug 2010, (Accessed on 05/10/2019).

\bibitem{ferreira2017certified}
J.~F. Ferreira, S.~A. Johnson, A.~Mendes, and P.~J. Brooke, ``Certified
  password quality - {A} case study using {C}oq and {L}inux pluggable
  authentication modules,'' in \emph{Integrated Formal Methods - 13th
  International Conference, {IFM} 2017, Turin, Italy, September 20-22, 2017,
  Proceedings}, 2017, pp. 407--421.

\end{thebibliography}

\end{document}